\documentclass[%
 aip,
 rsi,
 amsmath,amssymb,
preprint,%
]{revtex4-1}

\usepackage{graphicx}% Include figure files
\usepackage{dcolumn}% Align table columns on decimal point
\usepackage{bm}% bold math
\usepackage{siunitx}
\usepackage{hyperref}
\usepackage[utf8]{inputenc}
\usepackage[T1]{fontenc}
\usepackage{mathptmx}
\usepackage{etoolbox}
\usepackage{color}
\usepackage{soul}

\usepackage{ulem}
\newcommand\reduline{\bgroup\markoverwith
{\textcolor{red}{\rule[-0.5ex]{2pt}{0.8pt}}}\ULon}

\makeatletter
\def\@email#1#2{%
 \endgroup
 \patchcmd{\titleblock@produce}
  {\frontmatter@RRAPformat}
  {\frontmatter@RRAPformat{\produce@RRAP{*#1\href{mailto:#2}{#2}}}\frontmatter@RRAPformat}
  {}{}
}%
\makeatother
\begin{document}

\preprint{AIP/123-QED}

\title[Imaging based feedback cooling of a levitated nanoparticle]{Imaging based feedback cooling of a levitated nanoparticle}
% Force line breaks with \\
\author{Y. Minowa}
 \email[Author to whom correspondence may be addressed: ]{minowa@mp.es.osaka-u.ac.jp}
 \altaffiliation[Also at ]{JST, PRESTO, 4-1-8 Honcho, Kawaguchi, Saitama, Japan}%Lines break automatically or can be forced with \\
\author{K. Kato}%
\author{S. Ueno}%
\affiliation{ 
Graduate School of Engineering Science, Osaka University, 1-3, Machikaneyama-cho, Toyonaka, Osaka, 560-8531, Japan%\\This line break forced with \textbackslash\textbackslash
}%

\author{T. W. Penny}
\author{A. Pontin}
\affiliation{%
Department of Physics and Astronomy, University College London, Gower Street, London WC1E 6BT, United Kingdom%\\This line break forced% with \\
}%

\author{M. Ashida}%
\affiliation{ 
Graduate School of Engineering Science, Osaka University, 1-3, Machikaneyama-cho, Toyonaka, Osaka, 560-8531, Japan%\\This line break forced with \textbackslash\textbackslash
}%

\author{P. F. Barker}
\affiliation{%
Department of Physics and Astronomy, University College London, Gower Street, London WC1E 6BT, United Kingdom%\\This line break forced% with \\
}%

\date{\today}% It is always \today, today,
             %  but any date may be explicitly specified

\begin{abstract}
Imaging-based detection of the motion of the levitated nanoparticles complements a widely-used interferometric detection method, providing a precise and robust way to estimate the position of the particle. Here, we demonstrate a camera-based feedback cooling scheme for a charged nanoparticle levitated in a linear Paul trap. The nanoparticle levitated in vacuum was imaged by complementary metal-oxide semiconductor (CMOS) camera system. The images were processed in real-time with a microcontroller integrated with a CMOS image sensor. The phase-delayed position signal was fed-back to one of the trap electrodes which resulting in cooling by velocity damping. Our study provides a simple and versatile approach applicable for the control of low-frequency mechanical oscillators.
\end{abstract}

\maketitle

\section{Introduction}
High precision measurement and control of low frequency ($\lesssim$ kHz) mechanical oscillators are at the heart of recent advances in quantum/classical sensors based on levitated nano-oscillators\cite{pontinUltranarrowlinewidthLevitatedNanooscillator2020} and mesoscopic/macroscopic suspended oscillators\cite{catano-lopezHighMilligramScaleMonolithic2020, aguiarPresentFutureResonantMass2010}. In particular, nanoparticle oscillators levitated by electric\cite{conanglaMotionControlOptical2018a,daniaOpticalElectricalFeedback2021,pennyPerformanceLimitsFeedback2021a} or magnetic field\cite{slezakCoolingMotionSilica2018,gieselerSingleSpinMagnetomechanicsLevitated2020} are free from clamping losses and thermal damage due to optical absorption\cite{millenNanoscaleTemperatureMeasurements2014}. They are an attractive platform for quantum sensing, nanoscopic thermodynamics and tests of quantum physics\cite{pontinUltranarrowlinewidthLevitatedNanooscillator2020,millenOptomechanicsLevitatedParticles2020}. To achieve unprecedented sensitivities using these diverse family of levitation systems, different methods are needed to observe and control the motion of the levitated nano-oscillators precisely. A split detection scheme is widely used for observing the levitated particle motion,\cite{visscherConstructionMultiplebeamOptical1996,gittesInterferenceModelBackfocalplane1998,gieselerSubkelvinParametricFeedback2012,pennyPerformanceLimitsFeedback2021a,rahmanAnalyticalModelDetection2018} where the forward-scattered light from the particle is interferometrically detected with the unscattered reference light. Although the scheme provides very high spatio-temporal resolution, it is vulnerable to pointing fluctuations of the illuminating laser light\cite{bullierSuperresolutionImagingLow2019a} and has a relatively low dynamic range\cite{slezakCoolingMotionSilica2018}. In addition, if the particle is not well localised, the mean particle position can drift over long time periods preventing measurements of the particle dynamics over long periods \cite{bullierCharacterisationChargedParticle2020}.   

An alternative method has been proposed for low frequency levitated nano-oscillators \cite{bullierSuperresolutionImagingLow2019a} where the particle was imaged on to a high-speed CMOS image sensor and the acquired images were post-processed to determine the position of the particle. A diffraction limited image of the particle was fitted with a Gaussian function whose center corresponded to the estimated particle position. The process is similar to localization microscopy\cite{betzigImagingIntracellularFluorescent2006,betzigProposedMethodMolecular1995}, a family of super-resolution microscopies, providing sub-wavelength spatial resolution. The method ensures high dynamic range, and relatively flat noise characteristics down to DC\cite{bullierSuperresolutionImagingLow2019a}, and is thus suitable for the observation of the low frequency nanoparticle motion. However, real-time signal processing is required to control and cool the center-of-mass motion of the levitated nanoparticle with active feedback cooling\cite{pennyPerformanceLimitsFeedback2021a,conanglaMotionControlOptical2018a,daniaOpticalElectricalFeedback2021,gieselerSubkelvinParametricFeedback2012, gonzalez-ballesteroLevitodynamicsLevitationControl2021}.

In this paper, we report on feedback cooling of a charged nanoparticle levitated in a linear Paul trap. Here, the nanoparticle position signal was generated via the real-time processing of the nanoparticle images and was fed back to one of the trapping electrodes to apply a damping force proportional to the particle’s velocity (velocity damping scheme). All the image-processing procedures were executed on a microcontroller integrated with a CMOS image sensor, which ensures the high-speed and reliable image transfer between the CMOS image sensor and the microcontroller\footnote{Our code is available at \url{https://github.com/minowayosuke/high_fps_camera_feedback}}. Our proposed system provides a simple, flexible and low-cost way to observe and control the motion of the nano-oscillator and to study levitodynamics both classically and quantum mechanically\cite{gonzalez-ballesteroLevitodynamicsLevitationControl2021}.

\section{Charged nanoparticle levitated in a Paul trap}

\begin{figure}
\includegraphics[width=.95\columnwidth]{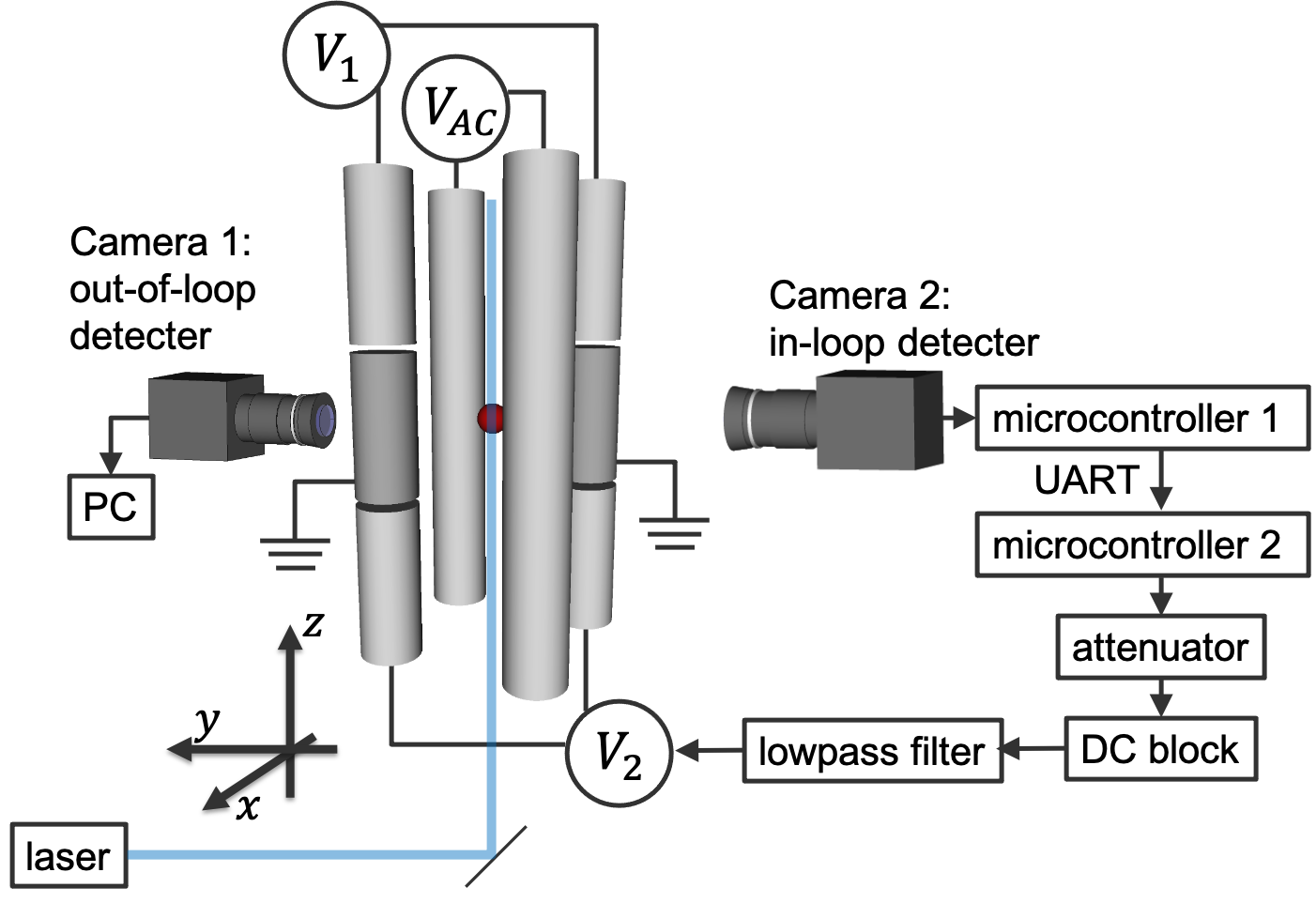}
\caption{\label{fig:setup} Experimental setup. A levitated nanoparticle was illuminated with a laser light beam and was imaged with a CMOS camera system (in-loop detector) The achieved images were processed with a microcontroller to obtain a voltage signal proportional to the particle position. The microcontroller shifts the phase of the signal along with another microcontroller to derive a signal corresponding to the particle velocity. A gain-adjusted and filtered signal was fed back to one of the end-cap electrodes for the feedback cooling.}
\end{figure}

Our linear Paul trap system consists of four cylindrical electrodes with a radius $R=\SI{4}{mm}$. A pair of diagonal electrodes provides an AC electric field to confine the nanoparticle's motion in the x-y plane while remaining confinement along the z-axis is provided by the two other segmented electrodes as shown in Fig. \ref{fig:setup}. The distance between the trap center and the AC electrodes is $r_0=\SI{6}{mm}$ and the axial distance between the trap center and the end-cap electrodes is $z_0=\SI{7}{mm}$. Before the injection of the nanoparticle, DC and AC voltages are typically set to $V_1=\SI{20}{V}$, $V_2=\SI{120}{V}$ and peak-to-peak voltage $V_{AC}=\SI{2000}{V}$, respectively. The AC drive frequency is kept at $\SI{600}{Hz}$. Silica nanoparticles ($d=\SI{450}{nm}$) were introduced into the trap via electrospray ionization\cite{kuhlickeNitrogenVacancyCenter2014} under ambient pressure. Normally, multiple particles are trapped simultaneously. Then, we can slightly adjust the DC end-cap voltage $V_1$ and $V_2$ to trap only one silica nanoparticle. After the successful trapping of a single nanoparticle, we set $V_1=V_2=\SI{10}{V}$ and $V_{AC}=\SI{600}{V}$. Then, we evacuate the whole chamber and initiate the following feedback cooling. A typical axial motional frequency of the trapped particle is $\omega_0/2\pi=20\sim 40$ Hz.

\section{Imaging based measurement of the particle position}
\begin{figure}
\includegraphics[width=.95\columnwidth]{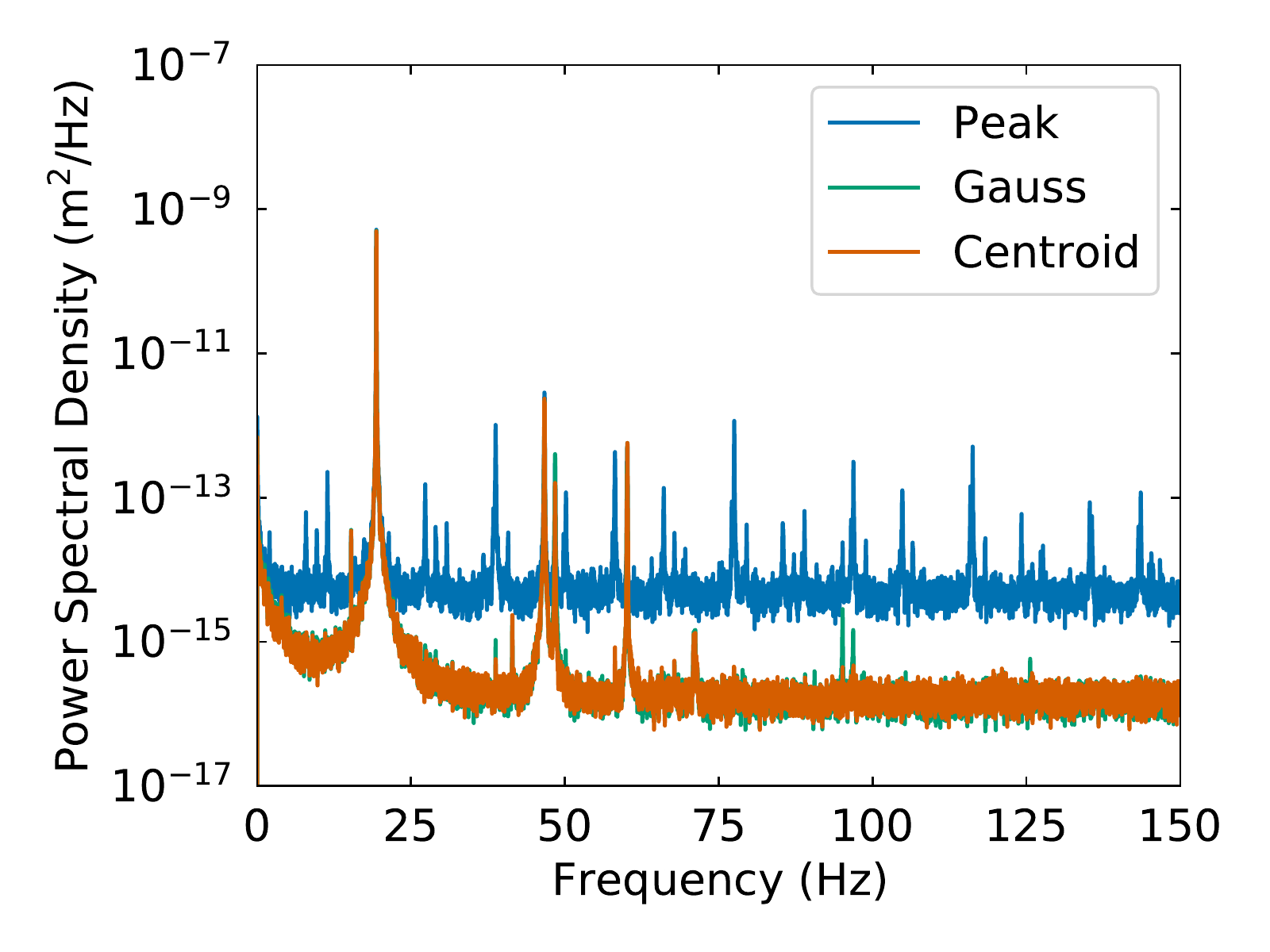}
\caption{\label{fig:centroid_demo} Comparison of different image processing schemes. Positional power spectral densities achieved at $4.3\times10^{-5}$ mbar using three different methods: peak detection (blue), Gaussian fitting (green) and centroid estimation (orange). The latter two curves nearly overlap.}
\end{figure}

The trapped nanoparticle is illuminated with a laser light beam and is imaged onto the CMOS camera sensors placed outside the vacuum chamber. Although the particle image itself is a simple indicator of the particle position, the image is broadened by the diffraction limit. Therefore, some image processing is needed to reconstruct the particle position. A standard procedure for the estimation is the two-dimensional Gaussian fit of the particle image, which is a technique used in super-resolution localization microscopy\cite{betzigProposedMethodMolecular1995,betzigImagingIntracellularFluorescent2006}. The Gaussian fit method provides us with a low-noise positional signal which is suitable for the precise characterization of the particle motion \cite{bullierCharacterisationChargedParticle2020,pontinUltranarrowlinewidthLevitatedNanooscillator2020,bullierSuperresolutionImagingLow2019a} However, the fitting procedure is problematic for the feedback cooling scheme because of its slow and non-deterministic processing time. To get the realtime signal corresponding to the particle position, a fast and deterministic method is required. Here, we chose the centroid estimation scheme, which is a simple and computationally-lightweight method. The method has been used in many fields in optics\cite{jiaMinimumVarianceUnbiased2010} and gives a precise position estimation comparable to the Gaussian fitting method\cite{cheezumQuantitativeComparisonAlgorithms2001}. 

For the comparison between  different image-processing schemes, we recorded the particle image sequences at $4.3\times10^{-5}$ mbar. The images were taken at 875.26 fps with camera 1 (see Fig. \ref{fig:setup}). The imaging based measurement enables us to easily calibrate the particle position. We mounted the CMOS camera sensor on a translation stage. By comparing the images before and after a certain amount of the lateral movement of the sensor, we can know the exact relation between the pixel number and the physical distance. We post-processed the recorded images with three different schemes to achieve the time trace of the particle position. In the first method, we identified the particle position by finding the brightest pixel position (peak detection scheme). The spatial resolution of the first method is limited by the pixel size of the sensor: \SI{5.35}{\micro m}. In the second method, each particle image was fitted with a two-dimensional Gaussian profile. The center of the Gaussian function provides us with the sub-pixel spatial resolution. In the third method, the particle position was calculated using the centroid estimation,
\begin{eqnarray}
{\bf r_{centroid}}=
\left(
\begin{array}{c}
x\\
z
\end{array}\right)\;
=C_\mathrm{cal}\dfrac{\sum_{i,j}{ {I_{i,j}}^p \left(
\begin{array}{c}
i\\
j
\end{array}\right)\;}}{\sum_{i,j}{I_{i,j}}^p},
\end{eqnarray}
where $i,j$ are integers representing the pixel position and $I_{i,j}$ is the signal count of the pixel $i,j$. Although, we chose $p=1$ here, we use $p=3$ to enhance the ratio of the signal to the background level later for the other camera system. Figure \ref{fig:centroid_demo} shows the positional power spectral densities (PSDs) achieved using the three different methods. It clearly shows a lower noise floor for the two sub-pixel methods, Gaussian fitting and centroid estimation than the peak detection scheme. The PSDs of the two sub-pixel methods nearly overlap indicating that we can implement the real-time feedback cooling using the centroid estimation without degrading the spatial resolution. We also measured the post-processing calculation time for each methods. For our system, the fastest is the peak detection scheme at 0.6 ms/frame and the centroid scheme at 0.9 ms/frame which is much faster than Gaussian fitting  at 5 ms/frame. Although the absolute calculation time depends on the implementation and the actual computational resources, the measured calculation times are good indicators of the relative time cost of each method.

\section{Velocity damping feedback cooling}
\begin{figure}
\includegraphics[width=.95\columnwidth]{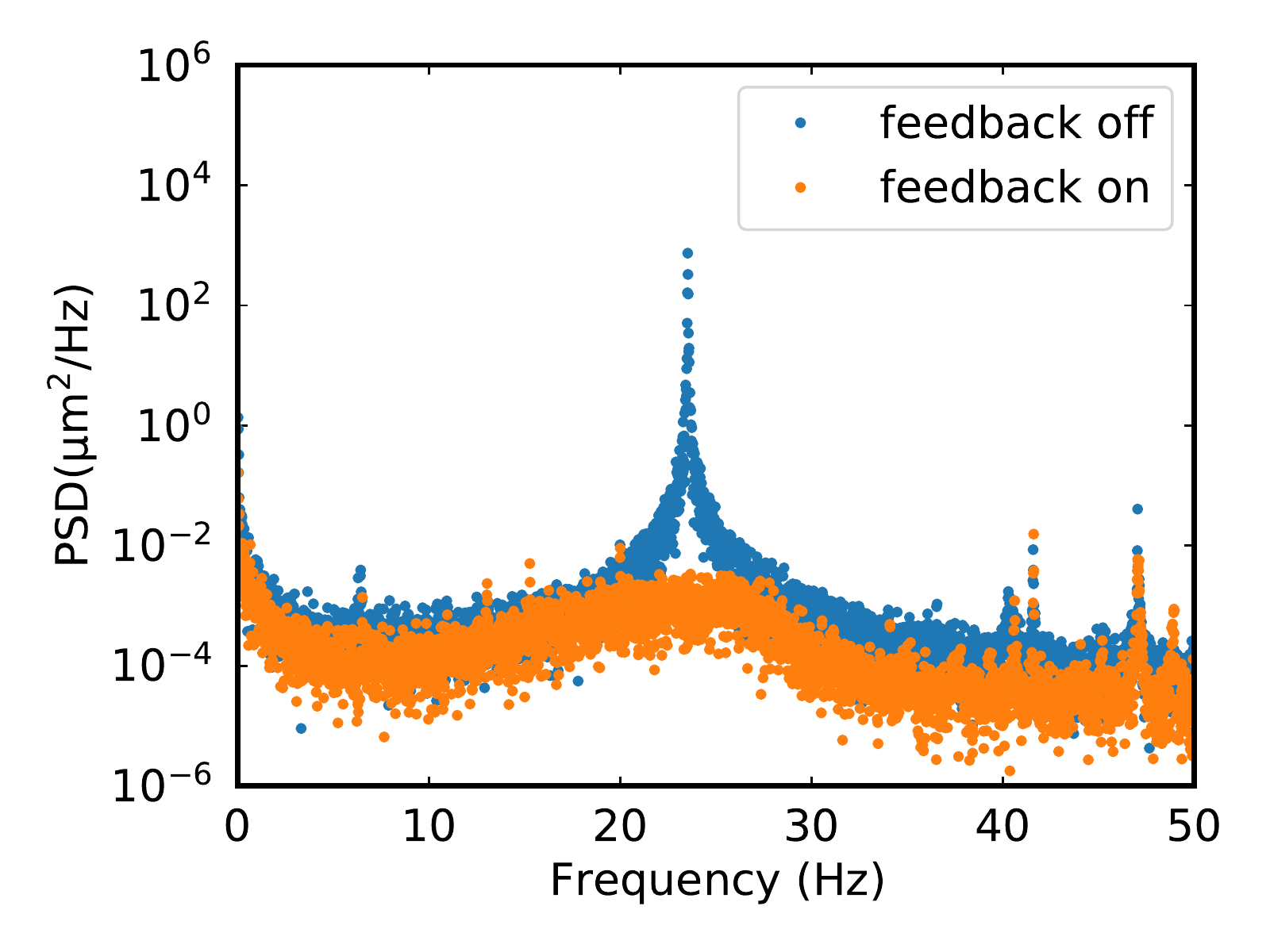}
\caption{\label{fig:typical_psd} Typical PSDs of the axial motion with $\omega_0/2\pi=23.5$ Hz. The data was taken with CMOS camera (out-of loop) at \SI{8e-5}{mbar} with and without feedback cooling.}
\end{figure}

Feedback cooling is implemented using an additional in-loop detector (camera 2, see Fig. \ref{fig:setup}). The camera sensor is integrated with a microcontroller, where the image processing is conducted in realtime. We fetch only a 20x30 pixel region of interest and process the image with the centroid scheme, which allows us to achieve $f_\mathrm{cam}=221$ fps. We chose velocity damping among possible feedback cooling schemes as it is simpler, more resilient and more efficient than parametric feedback cooling for our case\cite{pennyPerformanceLimitsFeedback2021a}.  Velocity damping cools the center-of-mass motion by applying an effective damping force proportional to the velocity of the nanoparticle. The feedback signal for velocity
damping can be derived by delaying the position signal by $\pi/2\omega_0$ instead of differentiating the signal which normally leads to a noisy signal.  It's easy to delay the signal by a multiple of $1/f_\mathrm{cam}$ by shifting the position signal over the processing iterations. However, the optimum delay $\pi/2\omega_0$ is generally not a multiple of the loop iteration time $1/f_\mathrm{cam}$. Therefore, we add another microcontroller for fine tuning of the delay. The digital signal protocol (UART) is used to transfer the data between the two microcontrollers to avoid any noise intrusion. The voltage signal proportional to the particle axial position is output from one of the digital to analogue converter (DAC) pin on the microcontroller 2. An appropriately attenuated and filtered feedback signal is applied to one of the end-cap electrodes, $V_2 = \SI{10}{V} +V_\mathrm{FB}$ resulting in cooling or heating depending on the amount of the delay. Fig.\ref{fig:typical_psd} shows typical PSDs with or without feedback derived from the nanoparticle position time trace; the data was taken with out-of-loop detector (camera 1). We can estimate the effective temperature of the center-of-mass motion by integrating the PSD\cite{millenOptomechanicsLevitatedParticles2020}. The area under the PSD curve is proportional to the effective temperature, 
\begin{equation*}
    \int_{0}^{\infty}S_{qq}(\omega)d\omega=\dfrac{k_B T_\mathrm{eff}}{m {\omega_\mathrm{cm}}^2},
\end{equation*}
where $S_{qq}(\omega)$ is the single-sided PSD, $T_\mathrm{eff}$ is the effective temperature, $m$ is the mass of the nanoparticle and ${\omega_\mathrm{cm}}$ is the angular frequency of the center of mass motion. We can calculate the effective temperature without knowing the mass by measuring the PSD at a higher pressure where the nanoparticle motion is in equilibrium with the surrounding air molecules motion at room temperature\cite{frangeskouPureNanodiamondsLevitated2018} so that the relevant quantity is simply the relative change of the peak area.

\begin{figure}
\includegraphics[width=.95\columnwidth]{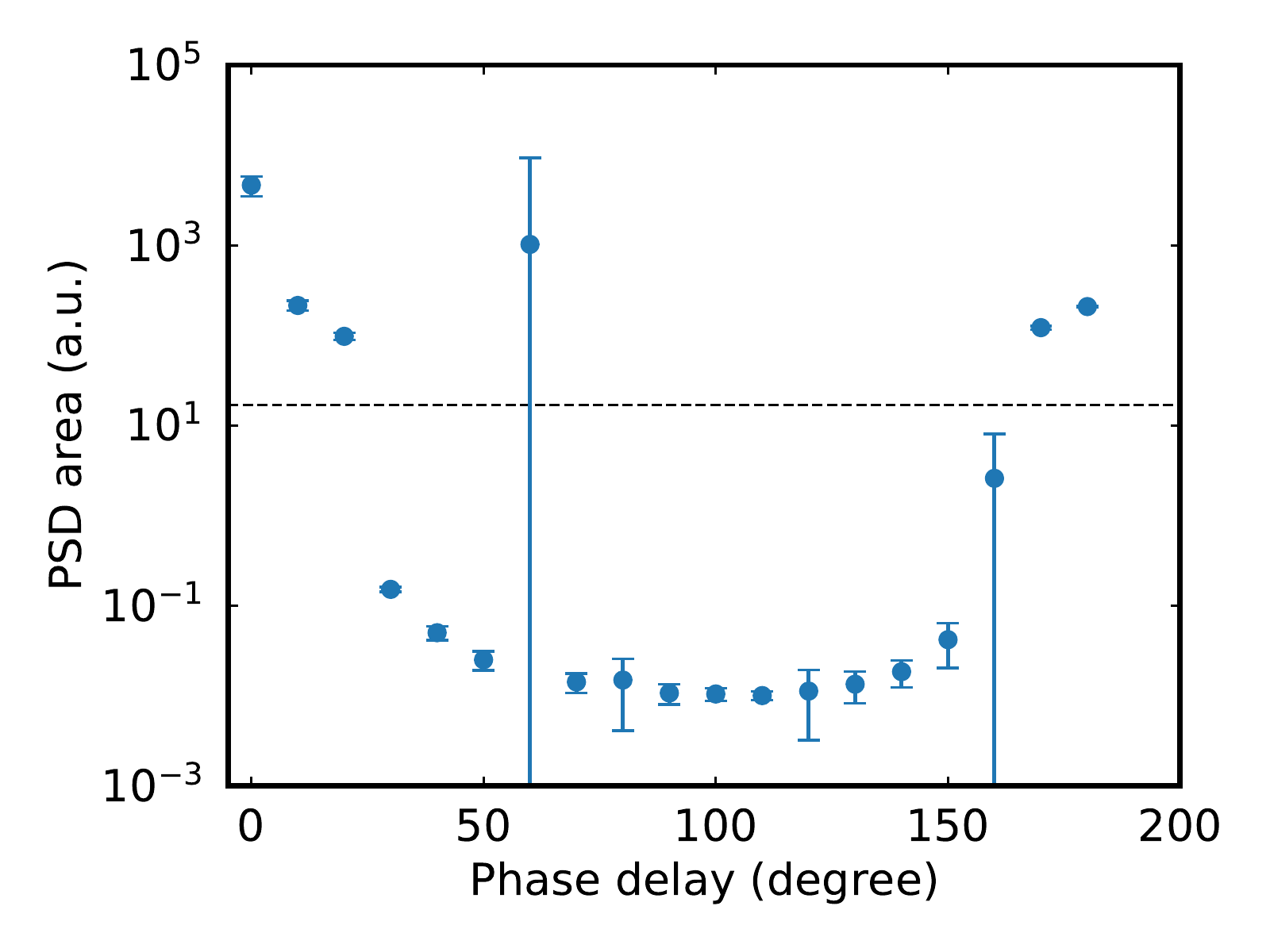}
\caption{\label{fig:phase} Phase delay dependence of the cooling at \SI{8.5e-5}{mbar}. Coarse and fine control of the phase delay was implemented with microcontrollers 1 and 2 respectively. Horizontal dashed line represents the PSD area without feedback at the same pressure.}
\end{figure}

We measured the phase delay and gain dependence of the cooling to find the optimum parameters. Fig.\ref{fig:phase} plots the PSD area as a function of the phase delay. It clearly shows the heating and cooling depending on the phase (the horizontal dashed line corresponds to the feedback-off data). We chose the optimum phase delay $\theta_0=\SI{110}{\degree}$ for further study. Note that our low pass filter which rejects unwanted signals adds an additional frequency dependent phase delay because the cut-off frequency is very close to the motional frequency $\omega_0$. The phase delay of the filter at $\omega_0$ is $\sim\SI{150}{\degree}$. Because the end-cap electrode position is opposite to the positive direction of the sensor coordinates this results in an additional \SI{180}{\degree} delay, the optimal total phase delay ($110+150+180=360+80$) is almost equal to $\SI{90}{\degree}$ as expected. Any small difference to this value is likely to result from the frequency dependent phase delay of the low-pass filter which has significant impact on the broadband signal of the cooled oscillator.

\begin{figure}
\includegraphics[width=.95\columnwidth]{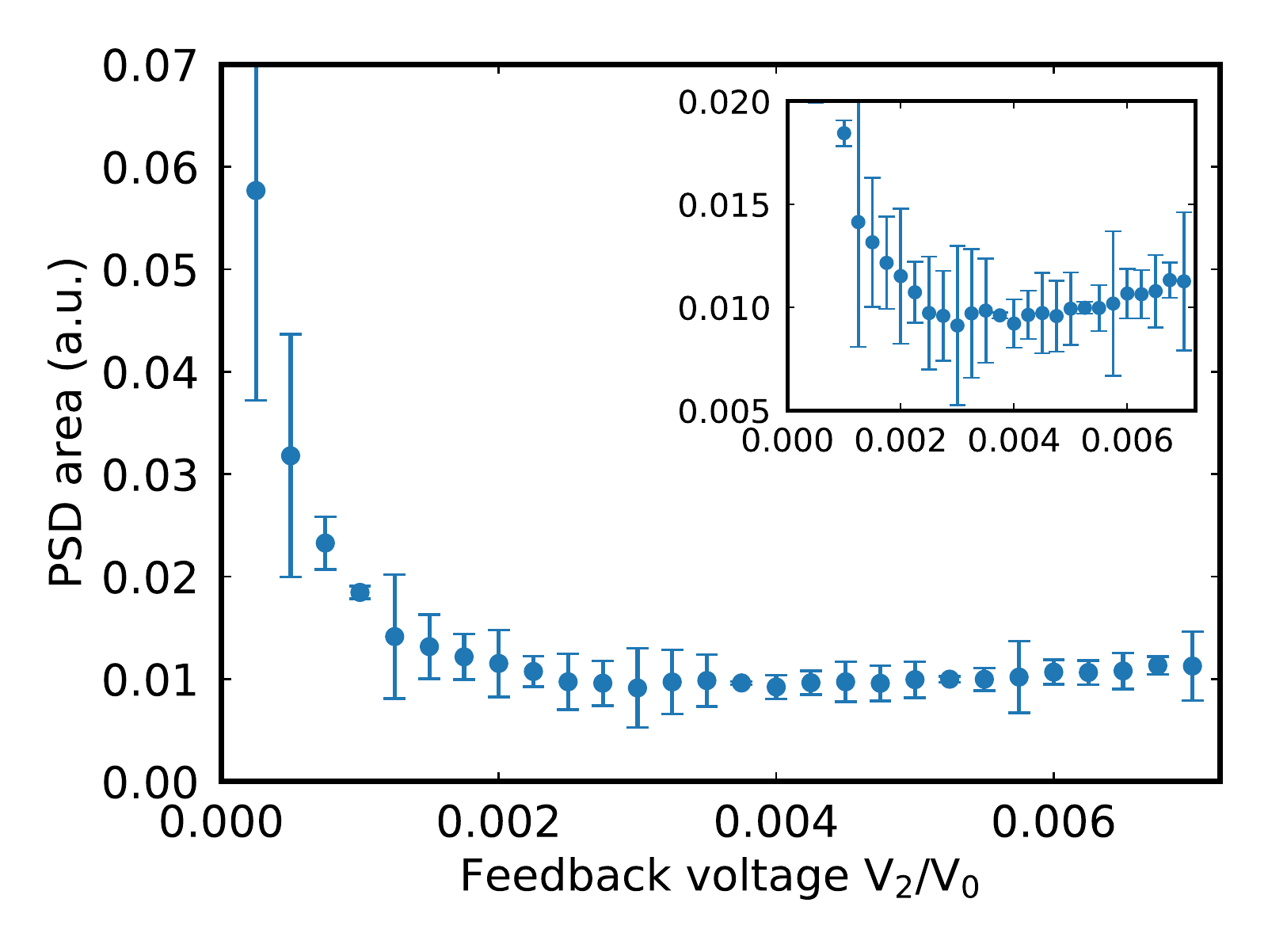}
\caption{\label{fig:gain} Feedback voltage dependence of the cooling at \SI{9e-5}{mbar}.}
\end{figure}

Figure \ref{fig:gain} shows the gain dependence of the cooling. When we increase the feedback voltage from zero, the cooling becomes more efficient reaching the lowest energy at $V_\mathrm{FB}/V_0=0.003$, where $V_0$ is the original voltage output from the DAC pin of the microcontroller 2. Further increase of the feedback voltage results in poorer performance. This may be due to the detection noise back action as previously reported\cite{pennyPerformanceLimitsFeedback2021a}. Any noise in the detection is fed back to the nanoparticle motion together with the appropriate velocity-damping feedback signal. The heating due to the detection noise surpasses the effect of the feedback cooling at higher feedback gain.

\begin{figure}
\includegraphics[width=.95\columnwidth]{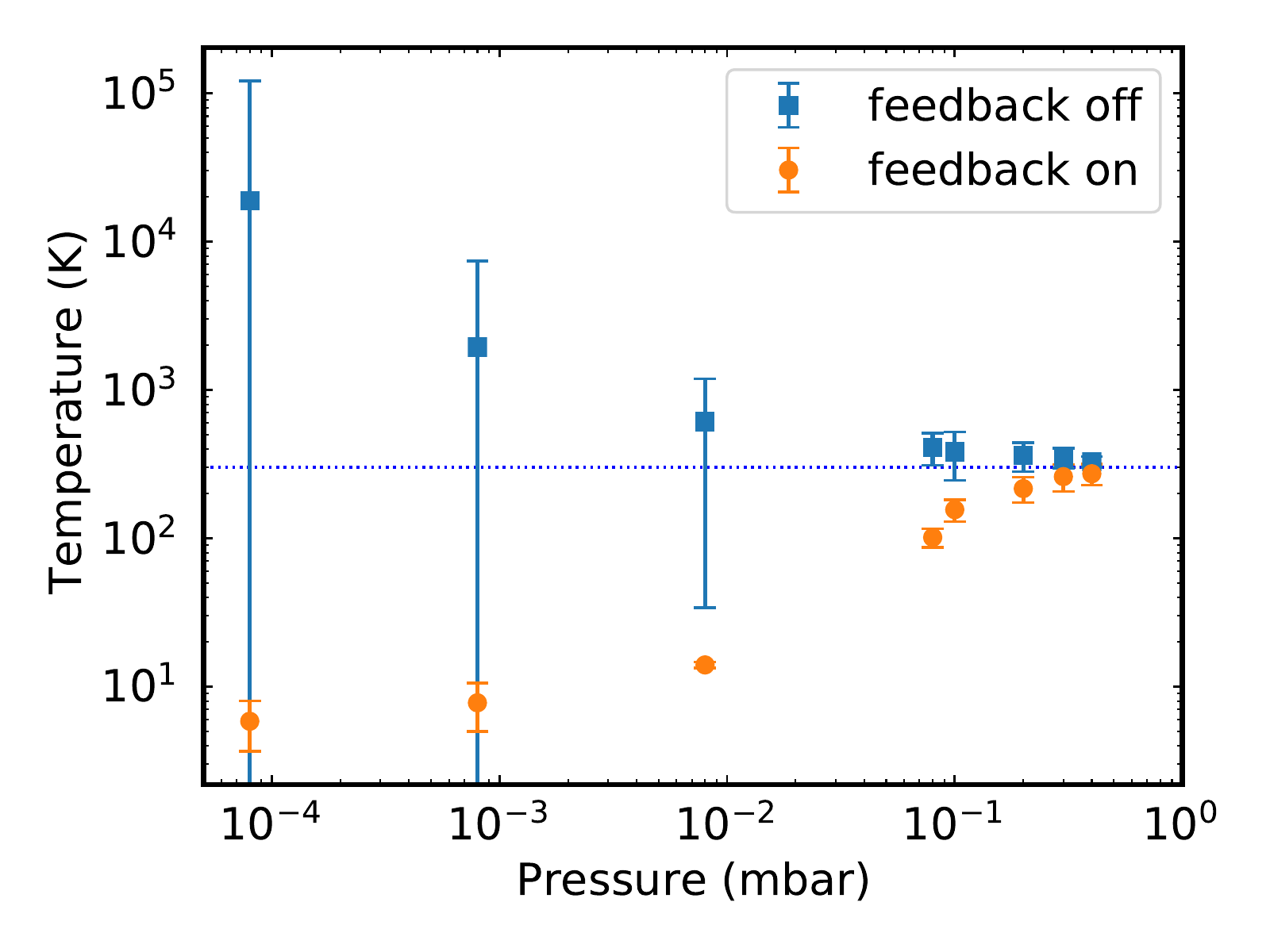}
\caption{\label{fig:pressure} Pressure dependence of the motional temperature with and without feedback. Feedback parameters are $\theta_0=\SI{110}{\degree}$ and $V_\mathrm{FB}/V_0=0.003$. The blue dotted line corresponds to the room temperature \SI{300}{K}.}
\end{figure}

We also measured the pressure dependence of the cooling to calibrate the energy of the center of mass motion. The PSD areas are proportional to the effective temperature of the center-of-mass motion as described above. We determined the proportional coefficient from the data taken at higher pressure where frequent collisions of the levitated nanoparticle with the gas molecules ensures thermal equilibrium at room temperature. As shown in Fig. \ref{fig:pressure}, the motional energies with and without feedback converge to a certain value at higher pressure, which should be equal to the room temperature $\SI{300}{K}$. We achieved the lowest temperature $\SI{5.8}{K}$ at $\SI{8e-5}{mbar}$. Importantly, the motional energy without feedback is increasing at lower pressure. This increase indicates the existence of other heating mechanisms which strongly affect the nanoparticle motion when the thermalization due to collision with air molecules is suppressed at lower pressures. Similar behavior has been reported before\cite{bullierCharacterisationChargedParticle2020, ranjitAttonewtonForceDetection2015a} and is caused by pressure-independent heating mechanisms including cross-coupling between different degrees of freedom\cite{pennyPerformanceLimitsFeedback2021a}, voltage noise in the circuit\cite{bullierCharacterisationChargedParticle2020} and seismic noise or pump vibration. Excess micromotion is another possible heating mechanism which can be reduced using some compensation electrodes\cite{daniaOpticalElectricalFeedback2021}. At lower pressure, the collisions between the nanoparticles and the gas molecules are less frequent. Here, the pressure-independent heating rate is more significant and the motional temperature becomes higher than the room temperature. Considering the ratio between the motional energies with/without feedback is $\gg10^3$, we can achieve much lower temperatures by eliminating the above mentioned heating mechanisms.

\section{Conclusions}
We have demonstrated the feedback cooling of a levitated nano-oscillator using imaging-based detection and feedback on the particle motion. The low frequency of these oscillators allows the use of a low-cost processing systems that integrates a CMOS sensor and a microcontroller. This allowed the implementation of different methods for digital signal processing of sequential images which may find application in other mechanical oscillator systems. One of the biggest advantages of the microcontroller-based detection system is a simple programming paradigm, which enables rapid-prototyping. We can use familiar programming languages such as C, C++, and python to implement digital signal processing and feedback cooling.  This feature contrasts with a Field-Programmable-Gate-Array (FPGA)-based system\cite{setterRealtimeKalmanFilter2018}, which would require a more complicated design flow using hardware description languages. The microcontroller-based detection system is best suited for feedback cooling and control of low-frequency ($\lesssim$ kHz) mechanical oscillators including levitated opto/electro/magnetomechanical systems particularly where long term drifts of the particle mean motion would preclude the use of typical difference detection methods. In addition, at least two axes of the motion could be captured on image and be used for feedback cooling and control. 

Further improvement and extension of the proposed scheme to higher frequency oscillators is possible by upgrading the microcontroller to implement more sophisticated digital signal processing\cite{setterRealtimeKalmanFilter2018,conanglaOptimalFeedbackCooling2019}. Such schemes could also be implemented using real time operation systems (RTOS) on a fast CPU, or using transfer of images directly to fast graphics processing unit (GPU) based without the requirement to be passed to the CPU\cite{liuRealTimeHumanDetection2021,smithFastSinglemoleculeLocalization2010}. Another interesting possibility is an execution of a loop protocol including the dynamic control of the harmonic potential shape\cite{weissLargeQuantumDelocalization2021}.

\begin{acknowledgments}
Y.M. acknowledge funding from the JSPS KAKENHI JP18KK0387 and JST PRESTO JPMJPR1909 and P.F.B. gratefully acknowledges support from the UK's EPSRC under grant No. EP/N031105/1 and EP/S000267/1, as well as the H2020-EU.1.2.1 TEQ project Grant agreement ID: 766900.
\end{acknowledgments}

\section*{Data Availability Statement}

The data that support the findings of this study are available from the corresponding author upon reasonable request.

\end{document}